\documentclass[aps,prl,twocolumn,showpacs,superscriptaddress,groupedaddress]{revtex4}  

\usepackage{graphicx}  
\usepackage{dcolumn}   
\usepackage{bm}        
\usepackage{amssymb}   
\usepackage{amsmath}
\usepackage{verbatim}
\begin{document}

\author{Junhui Shi}
\affiliation{Department of Chemistry, Princeton University, Princeton, NJ, USA}
\author{Suvi Ik\"{a}l\"{a}inen}
\affiliation{Department of Chemistry, University of Helsinki,
Finland }
\author{Juha Vaara}
\affiliation{Department of Physics, University of Oulu,
Finland}
\author{Michael  V.\ Romalis}
\affiliation{Department of Physics, Princeton University, Princeton, NJ USA}
\email{romalis@princeton.edu}

\title{Observation of optical chemical shift by precision nuclear spin optical rotation measurements and calculations}


\begin{abstract}
Nuclear spin optical rotation (NSOR) is a recently developed technique for detection of
nuclear magnetic resonance via rotation of light polarization,
instead of the usual long-range magnetic fields.  NSOR signals depend on hyperfine interactions
with virtual optical excitations, giving
new information about the nuclear chemical environment. We use a multi-pass optical cell to perform first precision measurements of NSOR signals for a
range of organic liquids and find clear distinction between proton signals for different compounds, in agreement with our earlier predictions. Detailed first principles quantum-mechanical NSOR calculations are found to be in good agreement with the measurements.
\end{abstract}

\maketitle

The effect of light on nuclear magnetic resonance (NMR) has been the subject of considerable interest, as it can be used to further increase the power of NMR spectroscopy \cite{Warren,Buckingham}. While NMR frequency shifts due to laser light turned out to be too small to be easily measured \cite{Warren1}, the inverse effect of optical rotation (OR) caused by nuclear spin polarization was observed in water and liquid $^{129}$Xe \cite{Savukov:2006hi}. The rotation of light polarization is similar to the Faraday effect caused by nuclear magnetization, but is enhanced by the hyperfine interaction between nuclear spins and virtual electronic excitations. Recent first-principles calculations of the nuclear spin OR (NSOR) predict different signals from chemically non-equivalent nuclei, opening the possibility of a new chemical analysis technique combining optical and NMR spectroscopy \cite{Ikalainen:2010vc}. However, the signal-to-noise ratio (SNR) of NSOR detection has been poor in the first experiments utilizing low-field CW NMR \cite{Savukov:2006hi} and in high-field pulsed
experiments \cite{Pagliero:2010uj,Meriles:2011}.  NSOR differences between the same
nuclei in different molecular positions have not been clearly observed \cite{Meriles:2011}.

Here we use a multi-pass optical cell \cite{Silver:2005tt, Li:2011} to measure proton NSOR with an SNR greater than 15 after 1000 seconds of integration and perform systematic NSOR measurements on organic liquids with absolute uncertainty of 5\%. Contrary to recent report \cite{Meriles:2011}, we find that the NSOR signals do not scale  with the Verdet constant $V$ of the liquid  since the hyperfine interaction between electrons and nuclei  is influenced by the chemical environment. The ratio of NSOR to Faraday rotation and to inductive $^{1}$H signals ranges by more than a factor of 2 for the simple chemicals studied. We also apply the recent
first-principles theoretical method~\cite{Ikalainen:2010vc} for calculation of
NSOR
and obtain results in good qualitative agreement with our measurements for both $^{1}$H and  $^{19}$F spins.

\begin{figure}[*h]
\begin{centering}
\includegraphics[scale=0.30]{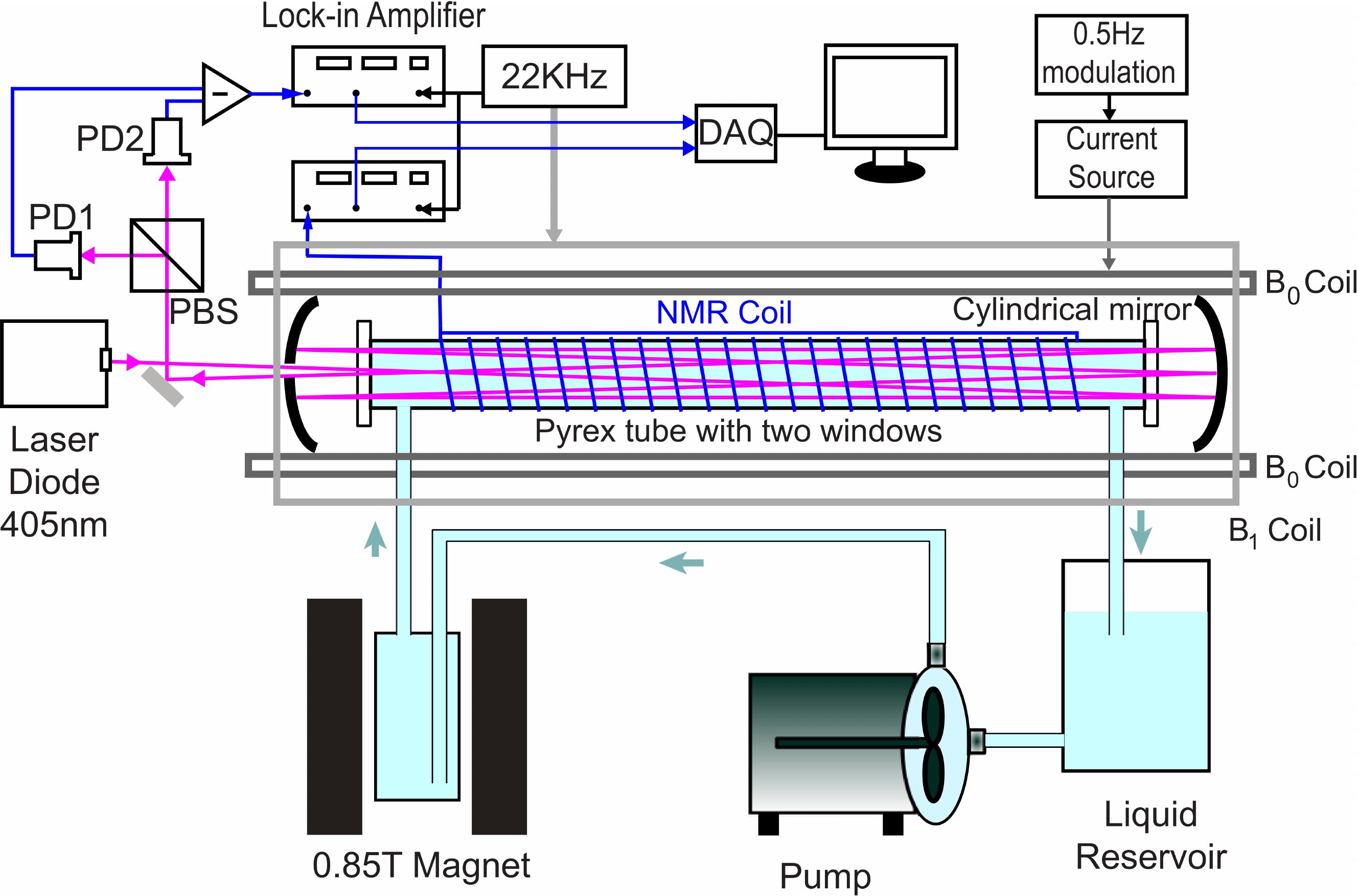}
\par\end{centering}
\caption{\label{fig:Detection Diagram} Apparatus for measurement of NSOR in organic
liquids.  The multi-pass cell consists
of two cylindrical mirrors with 40 cm separation and 50 cm radius of curvature, the sample tube is 22.5 cm long. }
\end{figure}

Optical cavities have been used in the past to amplify optical dispersion effects \cite{Zivattini,MerilesFaraday}, but multi-pass cells have a number of advantages \cite{Li:2011}. They do not require locking the frequency of the laser to the optical  cavity resonance or spatial mode matching of the laser beam to the cavity standing wave. They also do not require optimization of  mirror reflectivity to achieve maximum  power coupling into the cavity depending on losses.  Our multi-pass  cell consists of two cylindrical mirrors
with a small hole in one of the mirrors to let the laser beam enter and exit the cell (see Fig. 1) \cite{Silver:2005tt}. The number of passes is determined solely by the distance between the mirrors, their curvature, and the twist angle between their axes of curvature. The laser wavelength is chosen to be 405 nm, since NSOR is enhanced
at short wavelengths as $1/\lambda^2$.   To reduce optical losses in our cell, the sample tube end windows have an anti-reflection coating on the outside surfaces. While optical absorption length in very pure liquids can exceed 100 meters at 405 nm \cite{Waterabs}, it is very sensitive to impurities. We adjusted the multi-pass cell to have 14 passes for a total optical path length of 3.15 meters. The number of passes is determined by counting the number of beam spots on each mirror.  In the measurements on water the initial laser intensity of 8 mW was reduced to 0.4 mW after the multi-pass cell; the transmission was similar for other chemicals studied. Note that the increase in the photon shot noise by a factor of 4.5 due to light absorption is still less than the 14-fold increase in the OR signal, demonstrating an increase of the SNR with the multi-pass cell. The NSOR detection method  is based on a CW spin-lock
technique, first developed in Ref. \cite{Savukov:2006hi}. As shown in Fig.~\ref{fig:Detection Diagram}, the liquid
is circulated continuously by a pump from a reservoir to a permanent pre-polarizing magnet, and then to a
sample glass tube inside a uniform magnetic field $B_{0}=5 $ G.
An oscillating magnetic field is applied perpendicular to the $B_{0}$ field,
so the nuclear spins are adiabatically transferred to the frame rotating at the
NMR frequency as they enter the region of $B_0$ field. A  solenoidal pick-up coil wound around the
sample tube is used to measure the traditional NMR signal and determine the polarization for each flowing liquid~\cite{EPAPS}.

\begin{figure}[b]
\begin{centering}
\includegraphics[scale=0.34]{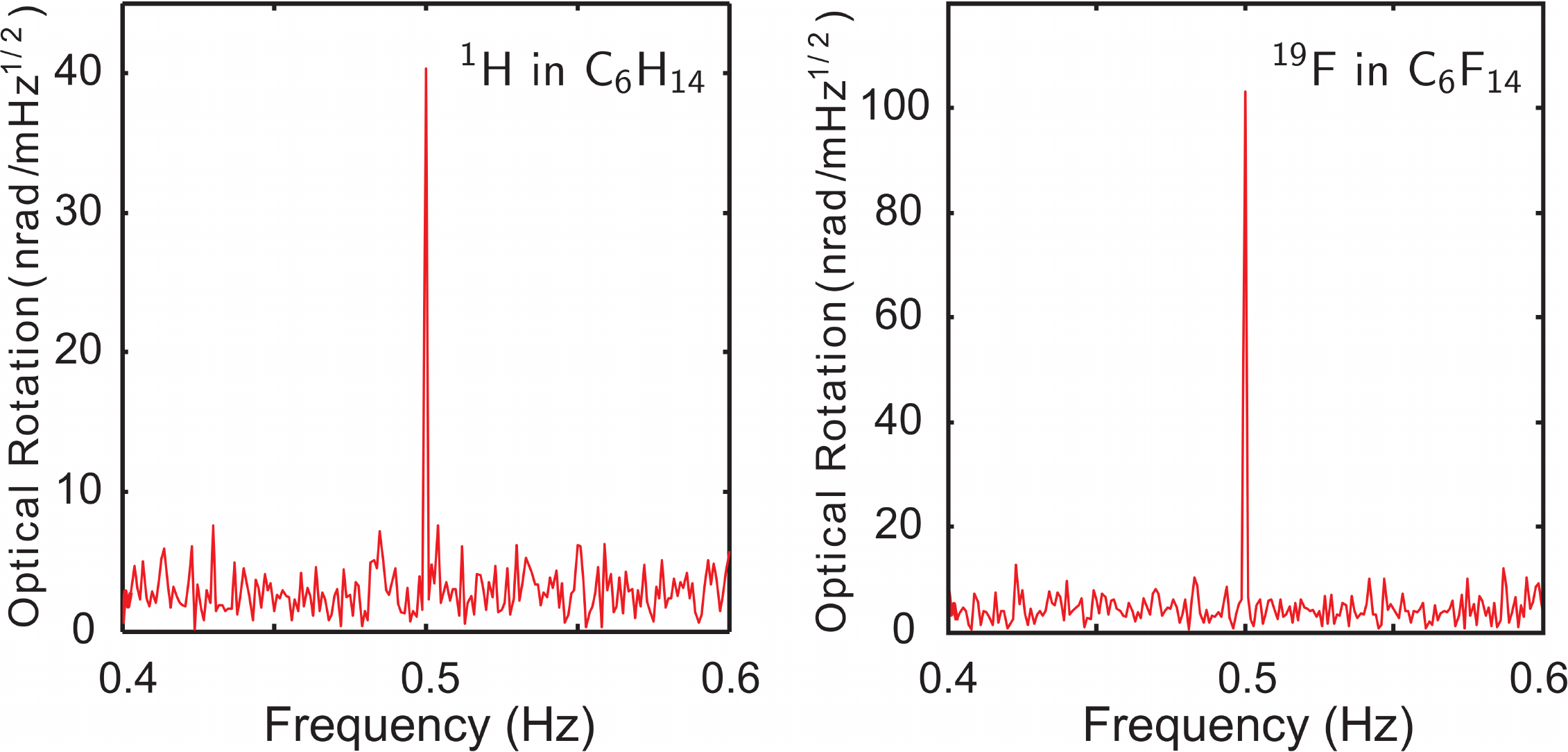}
\par\end{centering}

\caption{\label{fig:Optical-rotation-spectrum} Nuclear spin optical rotation signal of
$^{1}$H in $\mathrm{C_{6}H_{14}}$ and $^{19}$F in $\mathrm{C_{6}F_{14}}$
after 1000 seconds of integration. Since $B_{0}$ field is modulated on and off the resonance
at 0.5 Hz, the signal appears at this frequency.
The SNR is about 16 and 24 for $^{1}$H in $\mathrm{C_{6}H_{14}}$
and $^{19}$F in $\mathrm{C_{6}F_{14}}$ respectively. }
\end{figure}

The OR signal measured by a balanced polarimeter and the voltage across the pick-up coil are recorded by two lock-in amplifiers referenced to the NMR frequency.
In addition, we modulate the $B_{0}$ field on and off resonance at
0.5 Hz to distinguish NMR signals from any backgrounds. For static liquids the OR
noise is limited by photon shot noise, but it increases by about a factor of 2 during flow, likely due to small bubbles in the liquid.
The signal-to-noise ratio for water is typically about 15 after
one hour, while it is larger for other organic chemicals studied in this paper.
The NSOR signals of $^{1}$H in hexane and $^{19}$F in
perfluorohexane are shown in Fig.~\ref{fig:Optical-rotation-spectrum}.

\begin{figure}
\begin{centering}
\includegraphics[scale=0.40]{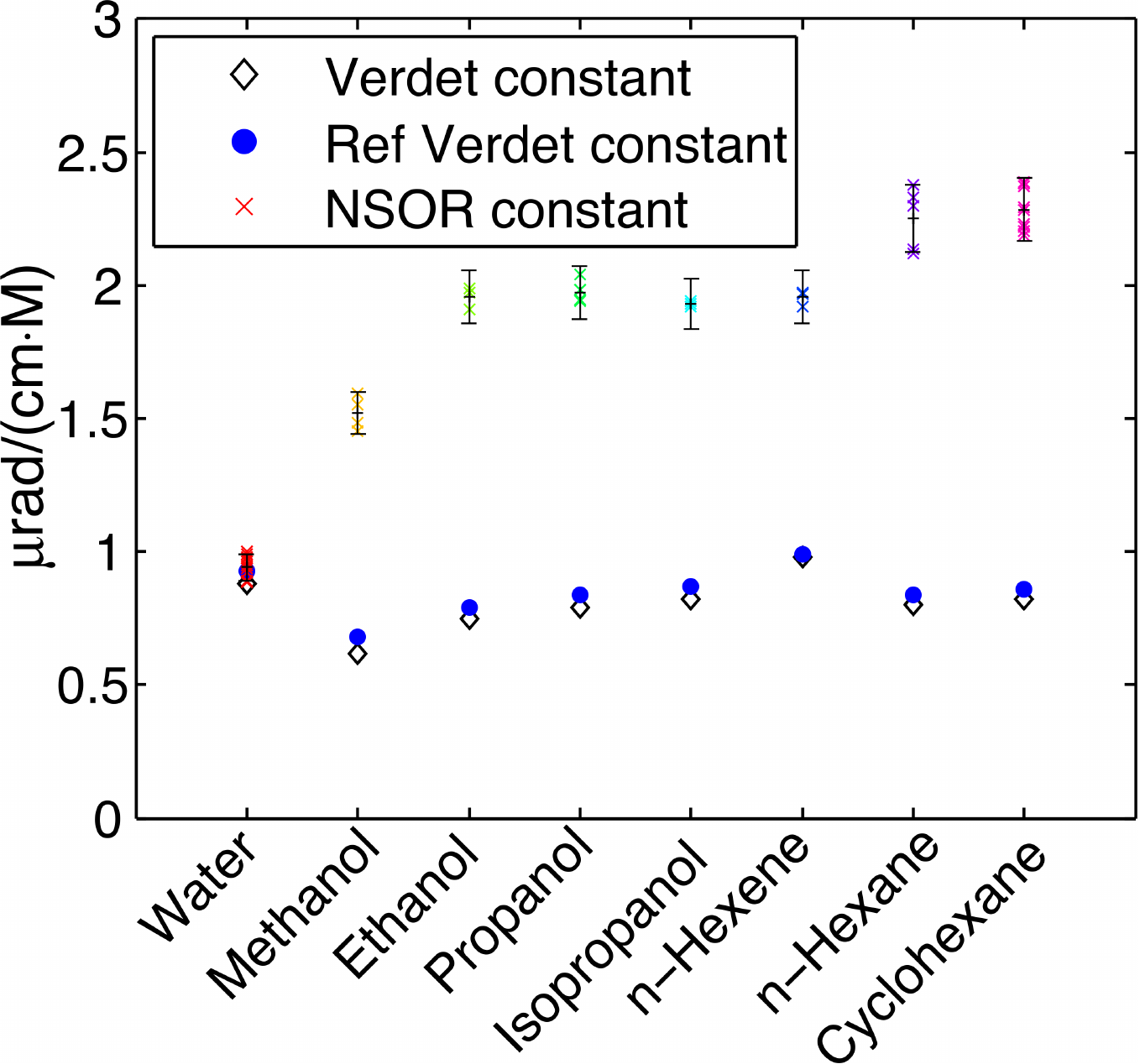}
\par\end{centering}

\centering{}\caption{\label{fig:NSOR and Verdet Graph}
$^{1}$H NSOR constants (cross
points and error bars), Faraday rotation Verdet constants (diamonds) from our
measurements and reference Verdet constants (blue dots).  All measurements of NSOR constants are shown for all chemicals to indicate the degree of experimental scatter. }
\end{figure}

In addition to NSOR, we also measured regular Faraday OR in all liquids
using the solenoid wound around the sample tube to create a known magnetic field.
The measured $V$ are used to calculate the
minimum NSOR signal due to the classical dipolar field produced by nuclear spins, given
by $\bm{B} = \mu_0 \bm{M}$ for a long sample tube.
Measurements of $V$ in a single-pass geometry revealed
that they are larger by  8.3\% compared to multi-pass geometry, likely due to polarization impurity caused
by multiple mirror reflections. We applied a $+8.3$\% correction to all OR data in the multi-pass geometry. The NSOR
were
measured several times
for each liquid, with periodic calibration by water NSOR measurements to check the long-term stability of the apparatus. In addition to the statistical
error we assign a systematic error of 5\% to each measurement that accounts for observed long-term changes in the signal amplitudes. We also find that our measured
$V$ are on average 5\% smaller than literature values at 405 nm \cite{Washburn:1926ts,Foehr,Villavarde}.  In Fig.~\ref{fig:NSOR and Verdet Graph} and \cite{EPAPS} we summarize the measured NSOR amplitudes and $V$, as well as the Verdet constants from the literature. The NSOR
data are scaled to 1~M concentration of fully polarized nuclei and $V$ values are multiplied by the classical magnetic field produced by these nuclei in a long cylinder.

For water the Verdet constant of Faraday OR
accounts for most of the NSOR signal, in agreement with \cite{Savukov:2006hi}, while in other chemicals  NSOR  is enhanced relative to Faraday
OR due to hyperfine interactions. The data for water, methanol and ethanol are generally
consistent with earlier first-principles theoretical calculations \cite{Ikalainen:2010vc}. The enhancement of  NSOR for hydrogens bound to carbon can be explained qualitatively by smaller electronegativity of C compared with O. This results in greater overlap of the electronic wavefunction with $^1$H in CH$_2$ and CH$_3$ groups, giving a larger hyperfine interaction.

We also measure $^{19}$F NSOR in perfluorohexane, which was first observed in \cite{Pagliero:2010uj} at high magnetic fields.  The NSOR signal after 1000~s is shown in Fig.~\ref{fig:Optical-rotation-spectrum}. The measured NSOR constant for $^{19}$F is a factor of 6 larger than in hydrocarbons, while $V$ is a factor of 3.5 smaller, consistent with earlier
data for $V$ in fluorocarbons \cite{Lagemann}. The enhancement of the NSOR signal by a factor of 57 relative to the OR due to the Faraday effect is partly due to stronger hyperfine interaction in heavier atoms \cite{Savukov:2006hi} and partly due to high electronegativity of fluorine.

The  OR angle $\Phi_{\rm{NSOR}}$ of a
beam of linearly polarized light arising due to nuclear spins can be calculated through the rotationally and ensemble-averaged antisymmetric polarizability~\cite{buc66,buc79,lu}. The
rotation per unit concentration $n$ of the polarized nuclei $N$ and per unit pass length $l$, $V_N=\Phi_{N{\rm SOR}}/n l$, is given by \cite{Ikalainen:2010vc},


\begin{align}
V_N =-&\frac{1}{2}\,\omega N_{\rm A}c\langle I_{N,Z}\rangle\,\frac{e^3\hbar}{m_e}\frac{\mu_0^2}{4\pi}\,\gamma_N \notag \\
 &\times \frac{1}{6}\sum_{\epsilon\tau\nu}\varepsilon_{\epsilon\tau\nu}{\rm Im}\langle\langle r_\epsilon;r_\tau,\frac{\ell_{N,\nu}}{r_N^3}\rangle\rangle_{\omega,0},
\label{eqn-nsor}
\end{align}
where $\omega$ is the frequency, $\langle I_{N,Z}\rangle$ is the average spin polarization,
$\gamma_{N}$ is the gyromagnetic ratio, and $\varepsilon_{\epsilon\tau\nu}$ is the Levi-Civita symbol. This expression is in terms of quadratic response theory~\cite{olsen}, involving time-dependent electric dipole interaction  with the
light beam taken to second order, and the static hyperfine interaction
\begin{equation} \label{pso}
H_{N}^{\rm{PSO}}=\frac{e \hbar}{m_e}\frac{\mu_0}{4\pi} \gamma_N \bm{I}_N \cdot \sum_i \frac{\bm{\ell}_{iN}}{r_{iN}^3}
\end{equation}
between the nuclear magnetic moment $\gamma_{N}\hbar\bm{I}_{N}$
and the
electrons. For heavy-atom systems such as liquid Xe,  relativistic formulation should be employed~\cite{ikalainen_xe}. In the present systems the nonrelativistic form (\ref{eqn-nsor}) is sufficient.

Eq.~(\ref{eqn-nsor}) does not include
the long-range magnetization field  discussed in Ref.~\cite{yao}. 
In a long cylindrical sample this
field equals $\bm{B}=\frac{1}{3}\,\mu_{0}\bm{M}$,
resulting in an additional Faraday rotation. Hence, a bulk correction $V_B$ given by
\begin{equation}
V_B=\frac{\Phi_{\rm B}}{n l} =\frac{1}{3}N_A \mu_0 \langle I_{N,Z}\rangle \hbar \gamma_N V,
\end{equation}
should be added to $V_N$ to be fully comparable to the experimental results.

Optimized geometries of the molecules were obtained with the Gaussian software~\cite{gaussian} at the B3LYP/aug-cc-pVTZ level, while the Dalton program~\cite{dalton} was used for
NSOR and $V$ at 405~nm. In the latter,  implementations of quadratic response functions were employed for the Hartree-Fock (HF), density-functional theory (DFT), and coupled-cluster (CC) methods from Refs.~\cite{hettema,salek,hattig}, respectively.  DFT was used to obtain results of predictive quality for larger molecules. Therefore, its performance was assessed through more accurate but also more time-consuming \emph{ab initio} CC singles and doubles (CCSD) calculations for
water, methanol, ethanol, propanol and isopropanol. The DFT functionals BHandHLYP(50\%), B3LYP (20\%), and BLYP (0\%) were used, where the percentages
denote the amount of exact HF
exchange admixture, which has been often seen to be the factor controlling DFT accuracy for hyperfine properties~\cite{vaara_pccp,Ikalainen:2010vc}.

Novel and compact sc.\ completeness-optimized (co) basis sets~\cite{manninen} were used to furnish near-basis-set limit results for
$V_N$,
which requires an accurate description of the electronic structure both at the nuclear sites and at the outskirts of the electron cloud. This is due to the involvement of both the magnetic hyperfine and electric dipole
operators.
The efficiency of
co
sets for magnetic properties has been verified in several
studies~\cite{manninen,laser_nmr,Ikalainen:2010vc,teemu_nsor,ikalainen_xe}. The co-2 set
(10$s$7$p$3$d$ primitive functions for
C--O;
10$s$7$p$3$d$ for H)
was developed in Ref.~\cite{laser_nmr} for laser-induced
$^{13}$C shifts in hydrocarbons. The carbon exponents~\cite{laser_nmr}
are used here also for oxygen.
Co-0
(C-O: 12$s$10$p$4$d$1$f$, H: 8$s$8$p$5$d$) was
generated in Ref.~\cite{teemu_nsor} for
basis-set-converged NSOR for
first-row main-group systems, as calibrated by $^{1}$H and $^{19}$FSOR calculations for the FH molecule.

To assess $V_{B}$, the
Verdet constants were
calculated
at the BHandHLYP/co-2 and B3LYP/co-2 levels for all molecules, as well as at the CCSD/co-2 level for water, methanol, ethanol, propanol, and isopropanol~\cite{EPAPS}.
$V$
computed with the 
co-2 basis were combined with $V_{N}$ obtained with co-0,
as the former property is not as sensitive to the basis-set quality as  NSOR.

\begin{figure}[tb]
\centering
\includegraphics[width=0.35\textwidth]{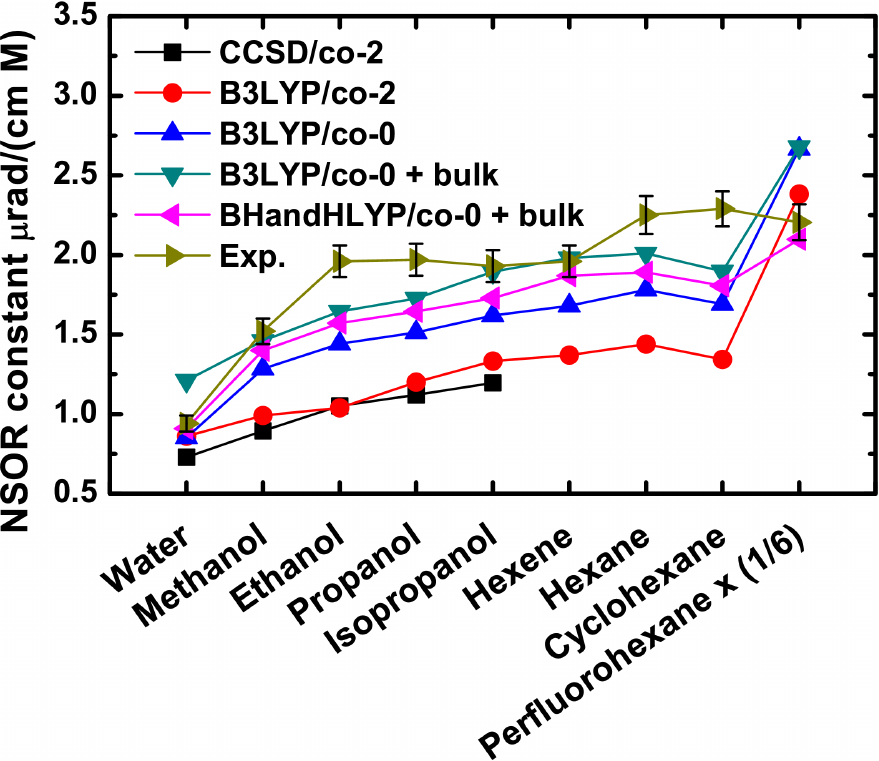}
\caption{$V_N$ for all molecules at the B3LYP/co-2 and B3LYP/co-0 levels of theory, as well as $V_N$ furnished with the bulk  correction for B3LYP/co-0 and BHandHLYP/co-0. CCSD/co-2 data is given for the smaller molecules.
Experimental values with error limits are also shown. Results for perfluorohexane are divided by a factor of six.}
\label{bulk_fig}
\end{figure}

Fig.~\ref{bulk_fig} and Table 1 in \cite{EPAPS} show the calculated NSOR.
Weighted averages over all $^{1}$H and $^{19}$F nuclei
of the molecules are reported. Furthermore,
tables of $V_{N}$ for the separate chemical groups are provided~\cite{EPAPS}. In most cases, the use of a basis set with higher quality leads to larger $V_N$. The only exception is water, where no systematic change is observed. Perfluorohexane shows a difference of  ca.~10\% between the two basis sets, while for the other molecules, the percentage ranges from 20--50\%. In all cases other than perfluorohexane, the calculated true NSOR [Eq.~(\ref{eqn-nsor})] is smaller than the experimental result.
Adding the bulk correction $V_B$ improves the agreement of BHandHLYP/co-0 and B3LYP/co-0 data with experiment (Figure~\ref{bulk_fig}), apart from the exaggerated B3LYP results for perfluorohexane and water.
However, the use of BHandHLYP/co-0 for water results in a good agreement with the measurements,  both due to the reduced true NSOR contribution as well as the more realistic $V$ obtained at the BHandHLYP level (Table~12 in \cite{EPAPS}). For the larger molecules, the experimental values are reproduced qualitatively. A detailed analysis would necessitate the incorporation of solvation and intramolecular dynamics effects via molecular dynamics simulations~\cite{teemu_nsor}. In the case of perfluorohexane, the known issues~\cite{vaara_pccp} of the present DFT functionals with the hyperfine properties of $^{19}$F may contribute to the observed overestimation.

Tables 3--11 in \cite{EPAPS} reveal that the largest NSOR occur in the CH$_{2}$ groups, while the hydroxyl groups display distinctly smaller values than either the methyl or methylene groups for all molecules. This supports the electronegativity argument for
the relatively small $V_{^{1}{\rm H}}$ in water ({\em vide supra}). In perfluorohexane, however, the CF$_3$-group feature a larger $V_{^{19}{\rm F}}$ than the CF$_2$ group. The different methylene groups in propanol and hexane, as well as the axial and equatorial hydrogens in cyclohexane give similar results. Alteration in the magnitude of $V_{^{1}{\rm H}}$ is observed for the CH$_{2}$ groups in hexene. The values for $^1$H in the methyl groups are rather similar for all the molecules, with isopropanol and hexane giving slightly larger NSOR than the other systems. The $V_{^{1}{\rm H}}$ appropriate to the hydroxyl groups differ between the molecules, with ethanol and propanol giving very small signals. In hexene, the {\em cis}-type hydrogen shows a larger rotation than the other protons situated next to the double bond.

In summary,
experimental NSOR signals do not scale with either the magnetic induction signal or the Verdet constant, but provide unique information about the nuclear chemical environment.
A simple
technique utilizing a multi-pass optical cell can be used to obtain signals with high SNR without
a superconducting magnet. Shorter multi-pass cells with larger number of passes~\cite{Li:2011}
can also be placed in superconducting magnets to obtain chemical shift information. Qualitative agreement with measured data
is obtained with first-principles calculations using DFT calibrated against \emph{ab initio} CCSD data. Hybrid DFT
with 20\%
or 50\%
exact exchange
is found to produce results closest to experiment.
The bulk magnetization correction described in Ref.~\cite{yao} is important. The $^1$HSOR is able to clearly distinguish hydroxyl group from  methyl and methylene groups. Future application of these techniques to more complicated molecules can provide unique new information about their confirmation and electronic wavefunctions.

Support has been
received
from
NSF grant
CHE-0750191 (JS and MR),
 the graduate school in
Computational Chemistry and Molecular Spectroscopy (SI),
and Academy of Finland (JV).
CSC (Espoo, Finland)
provided the computational facilities.

\bibliography{Paper1}

\end{document}